\documentstyle[12pt,amssymb,amscd]{article}

\def\AFOUR{%
\setlength{\textheight}{9.0in}%
\setlength{\textwidth}{5.75in}%
\setlength{\topmargin}{-0.375in}%
\hoffset=-.5in%
\renewcommand{\baselinestretch}{1.17}%
\setlength{\parskip}{6pt plus 2pt}%
}
\AFOUR
\def\car{\mathop{\square}}
\def\carre#1#2{\raise 2pt\hbox{$\scriptstyle #1$}\car_{#2}}

\makeatletter
\def\section{\@startsection {section}{1}{\z@}{-3.5ex plus -1ex minus
 -.2ex}{2.3ex plus .2ex}{\large\bf}}
\def\subsection{\@startsection{subsection}{2}{\z@}{-3.25ex plus -1ex minus
 -.2ex}{1.5ex plus .2ex}{\normalsize\bf}}
\makeatother
\makeatletter
\@addtoreset{equation}{section}

\makeatother
\newcommand{\nc}{\newcommand}
\newcommand{\rnc}{\renewcommand}
\nc{\be}{\begin{equation}}
\nc{\ee}{\end{equation}}
\nc{\bea}{\begin{eqnarray}}
\nc{\eea}{\end{eqnarray}}

\def\slash#1{\setbox0=\hbox{$#1$}#1\hskip-\wd0\hbox to\wd0{\hss\sl/\/\hss}}

\def\href#1#2{{#2}}

\rnc{\a}{\alpha}
\nc{\ab}{\bar{\a}}
\nc{\ap}{\a^{+}}
\nc{\abm}{\ab^{-}}
\rnc{\b}{\beta}
\nc{\bb}{\bar{\b}}
\nc{\bbp}{\bb_{\zb}^{+}}
\nc{\bm}{\b_{z}^{-}}
\nc{\oa}{\overline{\a}}
\nc{\ob}{\overline{\b}}
\rnc{\gg}{\gamma}
\rnc{\d}{\delta}
\nc{\f}{\phi}
\nc{\fb}{\bar{\phi}}
\nc{\vf}{\varphi}
\nc{\p}{\psi}

\rnc{\c}{\chi}
\nc{\la}{\lambda}
\nc{\m}{{\mathrm m}}
\nc{\n}{\nu}
\rnc{\o}{\omega}
\nc{\Om}{\Omega}
\rnc{\t}{\theta}
\nc{\eps}{\epsilon}
\rnc{\S}{\Sigma}
\nc{\F}{\Phi}
\nc{\trac}[2]{{\textstyle\frac{#1}{#2}}}
\nc{\ex}[1]{\mbox{e}^{\,\textstyle#1}}
\nc{\mat}[4]{\left(\begin{array}{cc}#1&#2\\#3&#4\end{array}\right)}
\nc{\som}[9]{\left(\begin{array}{ccc}#1&#2&#3\\#4&#5&#6\\#7&#8&#9%
\end{array}\right)}
\nc{\tr}{\mathop{\mbox{tr}}\nolimits}
\nc{\ad}{\mathop{\mbox{ad}}\nolimits}
\nc{\Tr}{\mathop{\mbox{Tr}}\nolimits}
\nc{\Det}{\mathop{\mbox{Det}}\nolimits}
\nc{\rk}{\mathop{\mbox{rk}}\nolimits}
\nc{\ra}{\rightarrow}
\nc{\Ra}{\Rightarrow}
\nc{\LRa}{\Leftrightarrow}
\nc{\ot}{\otimes}
\rnc{\ss}{\subset}
\nc{\nul}{\noindent\underline}
\nc{\non}{\nonumber\\}\nc{\subs}[1]{{\vspace*{0.5cm}}%
{\noindent\underline{#1}}{\addcontentsline{toc}{subsection}{#1}}%
{\vspace*{0.3cm}}}
\nc{\zb}{\bar{z}}
\rnc{\lg}{\frak{g}}
\nc{\lt}{\frak{t}}
\nc{\lk}{\frak{k}}
\nc{\lh}{\frak{h}}
\nc{\pik}{\Pi_{\lk}}
\nc{\pip}{\Pi_{+}}
\nc{\pim}{\Pi_{-}}
\nc{\pih}{\Pi_{\lh}}
\nc{\jz}{J_{z}}
\nc{\jzh}{\jz^{\lh}}
\nc{\jzp}{\jz^{+}}
\nc{\jzm}{\jz^{-}}
\nc{\del}{\partial}
\nc{\dz}{\del_{z}}
\nc{\dzb}{\del_{\bar{z}}}
\nc{\az}{A_{z}}
\nc{\azb}{A_{\bar{z}}}
\nc{\g}{g^{-1}}
\nc{\dw}{\Delta_{W}}
\nc{\Ad}{{\mbox{Ad}}}
\nc{\ks}{Ka\-za\-ma-\-Su\-zu\-ki}
\nc{\KS}{\ks}
\nc{\ksm}{\ks\ model}
\rnc{\AA}{{\Bbb A}}
\nc{\BB}{{\Bbb B}}
\nc{\CC}{{\Bbb C}}
\nc{\PP}{{\Bbb P}}
\nc{\cpm}{\CC\PP(m)}
\nc{\cpn}{\CC\PP(n)}
\nc{\cp}[1]{\CC\PP(#1)}
\nc{\gmn}{G(m,m+n)}
\nc{\gmnk}{\gmn_{k}}
\nc{\cO}{{\cal O}}
\nc{\bcO}{\bar{\cO}}
\nc{\bO}{\bar{O}}
\nc{\oQ}{\overline{Q}}
\nc{\ie}{{\it i.e.~}}
\nc{\eg}{{\it e.g.~}}
\begin{document}
\begin{flushright}
{}
\end{flushright}
\vspace*{0.3in}
\begin{center}
{\Large\bf D1D5 systems and AdS/CFT correspondences with 16 supercharges} \\
\vskip .3in
\makeatletter

\centerline{ Edi Gava\footnote{E-mail: gava@he.sissa.it}, \
Amine B. Hammou\footnote{E-mail: amine@sissa.it}, \ Jose F. Morales
\footnote{E-mail: morales@phys.uu.nl }, \
{\it and}
\ Kumar S. Narain \footnote{E-mail: narain@ictp.trieste.it} }
\smallskip\centerline{\it SISSA, INFN, Trieste, Italy$^{1,2}$}

\centerline{\it The Abdus Salam ICTP, Trieste, Italy$^{1,4}$}


\centerline{\it Spinoza Institute, Utrecht, The Netherlands $^3$}


\end{center}

{\bf Abstract:} We study the spectra of BPS excitations of D1D5
bound states in a class of free orbifolds/orientifolds of type IIB theory
and its dual descriptions in terms of chiral
primaries of the corresponding $AdS_3$ supergravities.

\vskip .10in

{\it Based on talks delivered by J.F. Morales at the 2001 RTN
Meetings: ``Conformal Symmetry and Strings'', Anttila, Sweeden and
``The quantum structure of spacetime and the geometric nature of
fundamental interactions", Corfu, Greece. }




$AdS/CFT$ correspondence \cite{maldacena} relates type IIB string
theory on $AdS_3\times S^3\times M$, with $M=T^4$ or $K3$, to
${\cal N}=(4,4)$ two-dimensional CFT's describing the 
infrared dynamics of D1D5 bound state systems. Tests
of this conjecture were performed in \cite{MS,DB1,DB2,mms}, where
multiplicities of BPS excitations of the CFT's
were shown to agree with those of chiral primary states
in the underlying supergravities. Although the supergravity
description is expected to be valid only for large values of the brane
charges, the correspondence was shown to work for all
$N=Q_1Q_5$, once a new additive quantum number, the {\it degree}
$d$, is introduced on the supergravity side \cite{DB2}. This is a
non-negative integer that 
allows to cut-off multiparticle states and implement the
exclusion principle \cite{MS}: one keeps only products of chiral
primaries whose total degree is $\leq N$.

In this paper we study the spectrum of chiral primary states and
their descendants in $CFT_2$/$AdS_3$ supergravity pairs, arising
from the D1D5 system in a class of freely acting $Z_2$
orbifolds/orientifolds of type IIB theory. Correspondingly, the near horizon
geometries are certain freely acting $Z_2$ orbifolds of
$AdS_3\times S^3\times T^4$.  The associated boundary CFTs have
been studied in great detail in \cite{ghmn}. The freely acting
$Z_2$ group generators are defined by accompanying the orbifold
and/or orientifold actions $\Omega$ (worldsheet parity), $I_4$
($Z_2$ reflection of $T^4$), $\Omega I_4$ with a shift
$\sigma_{p_6}$ along a circle transverse to the D1/D5 system, with
compact coordinate $X^6$. We refer to these theories as models
$I$, $II$ and $III$ respectively. 

In \cite{ghmn} the effective gauge theories associated to the type
IIB orbifold/orientifold D1D5 systems were argued to flow in the
infrared to CFTs locally equivalent to the one appearing for the
D1D5 system in type IIB theory on $T^4\times S^1$, but with
additional $Z_2$ global identifications induced by the orbifold
group actions \cite{ghmn}. The resulting target spaces in the
three models are of the form: 
\be 
{\cal M}_{\rm higgs}=\left(
R^3\times S^1\times T^4 \times (T^4)^N/S_N \right)/Z_2 \label{mh}
\ee
 The $Z_2$'s are generated by $(-)^{F_L}\, I_4^{\rm c.m.}$, $
I_4^{\rm c.m.}\,I_4^{\rm sp} $ and $(-)^{F_L}\, I_4^{\rm sp}$ for
the models $I$, $II$ and $III$ respectively, with $(-)^{F_L}$ the
left moving spacetime fermionic number, $I_4^{\rm c.m.}$ the
reflection of the first $T^4$ factor in (\ref{mh}) and  $I_4^{\rm
sp}$ the diagonal $Z_2$ reflection of the $N$ copies of $T^4$ in
the symmetric product part. Pure D1D5 states correspond to states
in the $Z_2$-untwisted sectors of (\ref{mh}). The resulting CFTs
are of type  ${\cal N}=(4,4)$  for models $II$ and ${\cal
N}=(4,0)$ for models $I$ and $III$, which involve the $\Omega$
world sheet parity projection\footnote{A closely related example of
${\cal N}=(4,0)$ D1D5 system where the $Z_2$ acts as a reflection 
of the transverse ${\bf R}^4$ accompanied with a longitudinal
shift have been recently studied in \cite{mt}. It would be
nice to apply the techniques developed in this paper to that system}.

The spectrum of charges and multiplicities of D1D5 BPS
excitations (right moving ground states $N_R=0$) was obtained from
the elliptic genus 
\bea \sum_N\, p^N\, \left( {\cal Z}^{(1)}_N+
{\cal Z}^{(g)}_N\right) &\equiv& \sum_N\, p^N\,{\rm Tr}_{{\cal
H}_N}\, \left( {1+g\over 2}\right)\, q^{L_0-c/24} \,y^{J^3_0}
\,\tilde{y}^{\bar{J}^3_0} \label{zh} 
\eea 
evaluated in the
CFT Hilbert spaces ${\cal H}_N$ defined by (\ref{mh}). $q=e^{2\pi
i\tau}$ describe the genus-one worldsheet modulus, $\bar{L}_0$,
$L_0$ are the Virasoro generators and $\bar{J}_0^3, J_0^3$ are
Cartan generators of an $SU(2)_R\times SU(2)_L$ current algebra to
which the sources $y$ and $\tilde{y}$ couple respectively. The
elliptic genera have been evaluated in \cite{ghmn} using
generalizations of the DMVV symmetric product formulas
\cite{dmvv}.

Following the general philosophy of Maldacena AdS/CFT
correspondences one can associated to these D1D5 CFT's
a dual description in terms of near horizon $AdS_3\times S^3$
supergravities.
 If we consider the radius $R_6$ of the circle along which the shift
is performed, very large, in such a way that the
space transverse to the D1D5 system is effectively $R^4$,
then the near horizon geometry will still be
$AdS_3\times S^3 \times T^4$.
However, in doing the KK reduction
the various modes will come with non-trivial $Z_2$ phases due to
the orbifold group actions ($\Omega$, $I_4$
and $\Omega I_4$ according to the model) in the way we will specify below.

The relevant $Z_2$-eigenvalues for 6-dimensional
massless fields of ${\cal N}=(2,2)$ supergravity, together with their
transformation properties under the little group $SO(4)$,
are displayed in the following table:\\\\
\begin{tabular}{lllll}
$I$ & $II$ & $III$ & \\
$\Omega$& $I_4$ &$\Omega I_4$& $bosons$ & $fermions$\\
$+$ &$ +$ &$+$& (2,2)+(0,2)+(2,0)+17(0,0) & 2(1,2) + 10(1,0) \\
$-$&$ +$&$-$& 4(0,2)+4(2,0)+8(0,0) & 2(1,2)+ 10(1,0)\\
$+$&$ -$&$-$& 8(1,1) & 2(2,1)+ 10(0,1) \\
$-$&$ -$&$+$& 8(1,1)& 2(2,1) + 10(0,1)\\
\end{tabular}\\\\\
{\it Table 2: SO(4) field content with $Z_2$ eigenvalues}\\\\
If we denote by $G_{-{1\over 2}}^i$,  $\tilde{G}_{-{1\over 2}}^i$
with $i=1,2$ the left and right moving lowering supersymmetry
charges in the AdS supergroup associated to the $AdS_3\times S^3\times T^4$
vacuum one can easily see that while $\tilde{G}_{-{1\over 2}}^i$
moves the states within each row, $G_{-{1\over 2}}^i$ moves vertically
between first two rows as well as the last two rows.

The spectrum of KK harmonics on $S^3$ can be determined
essentially by group theory \cite{SS,dkss,DB1}. The end result can
be written as \cite{ghmn} \be {\cal H}^A_{\rm
single~particle}=\oplus'_{m \geq 0} \, h_{r,s}\,({\bf m+r},{\bf
m+s})^{\epsilon_A(r,s)}_{m+1} \label{shortrs} \ee where $h_{r,s}$
denotes the Hodge numbers of the $T^4$ torus and \be ({\bf m},
{\bf m'})=\sum^2_{i,j=0} \pmatrix{i \cr 2}\pmatrix{j \cr 2} \,
(m-i,m'-j), \label{44} \ee collects states sitting in ${\cal
N}=(4,4)$ supermultiplets of the unorbifolded theory.
Supermultiplets $({\bf m},{\bf m'})^{\epsilon}_{d}$ are labeled by
the $SO(4)$ quantum numbers of the highest weight primary
$(m=2j,m'=2j')$ and are constructed by acting on this state with
the lowering operators $L_-,\bar{L}_-,J_0^{-},\bar{J}_0^-,
G_{-{1\over 2}}^i$, $\tilde{G}_{-{1\over 2}}^i$. The prime denotes
the omission of the term $m=r=s=0$ in the sum. The indices
$A=I,II,III$ label the model while  the subscript $m+1$ denotes the
degree. Finally $\epsilon_A(r,s)$ are the $Z_2$ eigenvalues of the
highest weight state inside the supermultiplet and are given by:
\be \epsilon^{I}(r,s) = (-1)^s,  ~~~~~~~~ \epsilon^{II}(r,s)
=(-1)^{r+s}, ~~~~~~~~\epsilon^{III}(r,s) = (-1)^r \label{phases}
\ee For models $I$ and $III$, although we have grouped the states
in terms of the original (4,4) multiplets, we have to remember
that $G^i_{-\frac{1}{2}}$ anticommutes with the $Z_2$'s, and
therefore the descendants that involve odd numbers of
$G^i_{-\frac{1}{2}}$'s will appear with an extra minus sign under
the $Z_2$ action.

Now we construct the multiparticle Hilbert space ${\cal H}_{\rm
multiparticle}$ and identify the finite $N$ CFT Hilbert space with
the subset of states in ${\cal H}_{\rm multiparticle}$ that have
degree less than or equal to $N$. To this end it is convenient to
introduce a parameter $p$ that keeps track of the degree
$d={\sum_i \, (m_i+1)}$ of multi-particle states.
 Carrying out the trace over all
(chiral,chiral) primaries above one is left with the result 
\be
{\cal Z}^g(p,q,y,\tilde{y})=\prod_{m=0}^{\infty}\prod_{r,s=0}^2
(1-\epsilon^A(r,s)p^{m+1}y^{m+r}\tilde{y}^{m+s})^{-(-)^{r+s}h_{r,s}}+O(q)
\label{eg}
\ee The factor $\frac{1}{1-p}$ ($m=r=s=0$) 
previously missing in (\ref{shortrs}) takes
into account the fact that finite $N$ CFT Hilbert space is
identified with states in ${\cal H}_{\rm multiparticle}$  that
have degree up to $N$ and not just $N$. It is easy to see that
the elliptic genera (\ref{eg})
exactly reproduce the CFT results (formulas (5.13) in \cite{ghmn0} 
for the three models after spectral flow from Ramond to NS sector
\cite{ghmn}.

One can now, following de Boer, check the correspondence beyond
the (c,c) primaries. The idea is to construct the finite $N$
elliptic genus which is obtained by taking the trace over states
of the form chiral on the right-moving sector and any state on the
left-moving sector, and setting $\tilde{y} = \bar{q}^{-1/2}$. The
comparison of the states can, of course, only be made for
dimensions much less than $N$, since otherwise gravity
approximation would break down. In \cite{DB1} de Boer showed that
for the K3 case the matching of the states goes all the way up to
left dimension equal to $(N+1)/4$. This is exactly the bound at
which black hole is expected to form. In the right-moving sector
arbitrary chiral states are allowed, and they have a bound on the
dimension which is of order $N/2$.

We will restrict our elliptic genus computation to
models $II$ and $III$ since the elliptic genus of 
model $I$ vanishes for $\tilde{y} = \bar{q}^{-1/2}$.
Here one can repeat the
analysis of \cite{mms} and take 2 derivatives with respect to $\tilde{y}$ 
before setting it to $\bar{q}^{-1/2}$. In this case however we do not get any
information; in fact for states satisfying the bound $h \leq (N+1)/4$ only
the ground state contributes as shown in \cite{mms}. 

The elliptic genus for the multi-particle states is then \be
Z_{\rm multiparticle}^A(p,q,y) =\prod_{n,m,\ell}\left[ \frac{1+p^n
q^m y^{\ell}}{1-p^n q^m y^{\ell}} \right]^{c^{A+}_{\rm
sgr}(n,m,\ell)}. \label{egsgrmp} \ee where $c^{A+}_{\rm
sgr}(n,m,\ell)$'s are the coefficients in the expansion of \bea
Z^A(p,q,y) =&&\frac{1}{2} \sum_{m,r,s} \sum_{t=0}^{min(2,m+s)}
{\sum_{k=0}^{\infty}}' d(r)d(s)d^A(t)\epsilon^A(r,s)
p^{m+1}q^{\frac{m+s+t}{2}+k}\nonumber\\ &&\times
\sum_{j=0}^{m+s-t} y^{m+s-t-2j} \label{egsgrsp1} \eea where we
have used the fact that $(-1)^{r+s} h_{r,s} = d(r).d(s)$ with
$d(0)=d(2)=1$ and $d(1)=-2$. The sum over $m,r,s$ takes into
account all (c,c) primaries $(m+r,m+s)$. We have also included
here the $m=r=s=0$, which is not a (c,c) primary in supergravity,
to take into account the redefined $Z$'s including a shift by
$p/2$. The sum over $k$ takes into account the descendents coming
from applying $L_{-1}$ and the primed sum over $k$  means that for
$m+s=0$ there is only one term in the sum, namely $k=0$, and for
$m+s \neq 0$ the sum is over all the non-negative integers. This
is due to the fact that for $m+s=0$ we have the left ground state
and $L_{-1}$ annihilates this state. The sum over $j$ takes care
of the descendents coming from applying $J_-$ and finally sum over
$t$ takes into account the descendents coming from applying
$G^i_{-\frac{1}{2}}$. The upper bound on this sum means that if
$m+s$ is less than 2, then we can only apply a maximum of $m+s$
$G^i_{-\frac{1}{2}}$'s to the chiral primary. $d^A(t)$ takes into
account the multiplicities of these descendents, together with the
$Z_2$ actions. Since in model $II$, the $Z_2$ commutes with (4,4)
supersymmetry, $d^{II}(t) =d(t)$. On the other hand for models $I$
and $III$, the $Z_2$'s anticommute with $G^i_{-\frac{1}{2}}$'s and
therefore $d^I(t)=d^{III}(t)=d(t)$ for $t$ even and
$d^I(1)=d^{III}(1)= -d(1)=2$. Note that since $\epsilon^I(r,s) =
(-1)^s$ the sum over $r$ yields zero on the right hand side as
expected for model $I$. For models $II$ and $III$, since
$\epsilon(r,s)$ contains $(-1)^r$, the summation over $r$ gives a
factor of 4.

The same applies also to the CFT side. but now with $c_{\rm
cft}^{A+}(m,\ell)$ given by the expansion coefficients 
of the partition function ${\cal Z}^g_{N=1}$
over the (4,4) CFT with target space $T^4$
in zero momentum and winding sectors. The $Z_2$ orbifold actions
are given by $g^{II}=I_4$ and
$g^{III}=(-1)^{F_L}I_4$. It follows that 
\begin{eqnarray}
\sum_{m,\ell}c_{\rm cft}^{II+}(m,\ell)q^m y^{\ell} &=& 8 \left[\frac{\theta_2(q,z)}
{\theta_2(q,0)}\right]^2 \nonumber\\
\sum_{m,\ell}c_{\rm cft}^{III+}(m,\ell)q^m y^{\ell} &=&- 8 \left[\frac{\theta_1(q,z)}
{\theta_2(q,0)}\right]^2 
\end{eqnarray}
where $y=e^{2\pi i z}$.

The matching of CFT states and supergravity states for dimensions
less than $N/4$ implies the following relations \bea \sum_nc_{\rm
cft}^{A+}(4mn-n^2-\ell^2)&=&\sum_n c_{\rm sgr}^{A+}(m,n,\ell)
\nonumber\\ \sum_n nc_{\rm cft}^{A+}(4mn-n^2-\ell^2)&=&\sum_n
nc_{\rm sgr}^{A+} (m,n,\ell) \label{matching} \eea between the
supergravity and CFT expansion coefficients. A detailed evaluation
of the quantities entering in both sides of (\ref{matching}) for
model II and III can be found in
 \cite{ghmn}.
 The results show a complete agreement between
supergravity and CFT computations for model II in the expected
range of validity. For model $III$ however, we find a discrepancy
for states with $\ell=0$ and $m > 0$. This mismatch at the
3-charge level have been already observed in \cite{ghmn0}, when
CFT counting formulas were tested against U-duality.
 More precisely, consistency with U-duality requires
that counting formula should be invariant under the simultaneous
exchanged of model $II$ and $III$ and the $D_1$ and $p_1$ charges.
This implies that the elliptic genus for model $III$ should be the
same as model $II$ with $p$ and $q$ exchanged. One can wonder
whether the supergravity result in the model III agrees with the
CFT counting formula in model $II$ after exchanging $p$ and $q$.
 This turns out to be the case (see \cite{ghmn} for details),
 allowing us to test the ${\cal N}=(4,4)$ CFT in
the regimes of small and large conformal dimensions by two
different supergravity duals.

\vskip 0.5in
{\bf Acknowledgements}

 This project is
supported in part by EEC under TMR contracts ERBFMRX-CT96-0090,
HPRN-CT-2000-00148 , HPRN-CT-2000-00122 and the INTAS project
991590.

\rnc{\Large}{\normalsize}

\end{document}